\documentclass[
 reprint,
 amsmath,amssymb,
 APS,
prb,
]{revtex4-2}

\usepackage{graphicx}% Include figure files
\usepackage{dcolumn}% Align table columns on decimal point
\usepackage{bm}% bold math
\usepackage{xcolor}
\usepackage{amsmath}
\usepackage{ulem}
\usepackage{soul}
\usepackage{hyperref}% add hypertext capabilities

\newcommand{\beginsupplement}{%
 	\setcounter{page}{1}
 	\renewcommand{\thepage}{S\arabic{page}}%
 	\setcounter{figure}{0}
 	\renewcommand{\thefigure}{S\arabic{figure}}%
 }

\begin{document}

\title{Static and dynamic analysis of auxetic three-dimensional curved\\ metamaterials in both axial and circumferential directions}

\author{Mohamed Roshdy, Osama R. Bilal}
\email{osama.bilal@uconn.edu}
 \affiliation{School of mechanical, aerospace, and manufacturing engineering, University of Connecticut, Storrs, CT, 06269, USA}

\date{\today}

\begin{abstract}
Metamaterials can enable unique mechanical properties based on their geometry rather than their chemical composition. Such properties can go beyond what is possible using conventional materials. Most of the existing literature consider metamaterials in Cartesian coordinates with zero curvature. However, realistic utilization of meta-structures is highly likely to involve a degree of curvature. In this paper, we study \textcolor{black}{both} the effective static and dynamic properties of metamaterials in the presence of curvature. \textcolor{black}{To capture the effect of curvature on the static behavior of our metamaterial, we calculate the effective Poisson's ratio of the metamaterial in the presence of curvature}. We conduct our analysis on three-dimensional metamaterials with varying effective Poisson's ratio. \textcolor{black}{We observe a significant change in the values of the effective Poisson's ratio of the metamaterial duo to curvature}. To capture the effect of curvature on the dynamics of our metamaterials, we calculate dispersion curves of curved metamaterial at different circumferential directions. We show both numerically and experimentally the change of the dynamic behavior of auxetic metamaterial from attenuation to transmission and vice-versa due to curvature. Our findings underscore the importance of curvature in both static and dynamic analysis of metamaterial design and could provide the means to guide practical implementations of metamaterials for functional use.

\end{abstract}

\maketitle

\section{Introduction }

\begin{figure*}
\centering
\includegraphics[width= \textwidth]{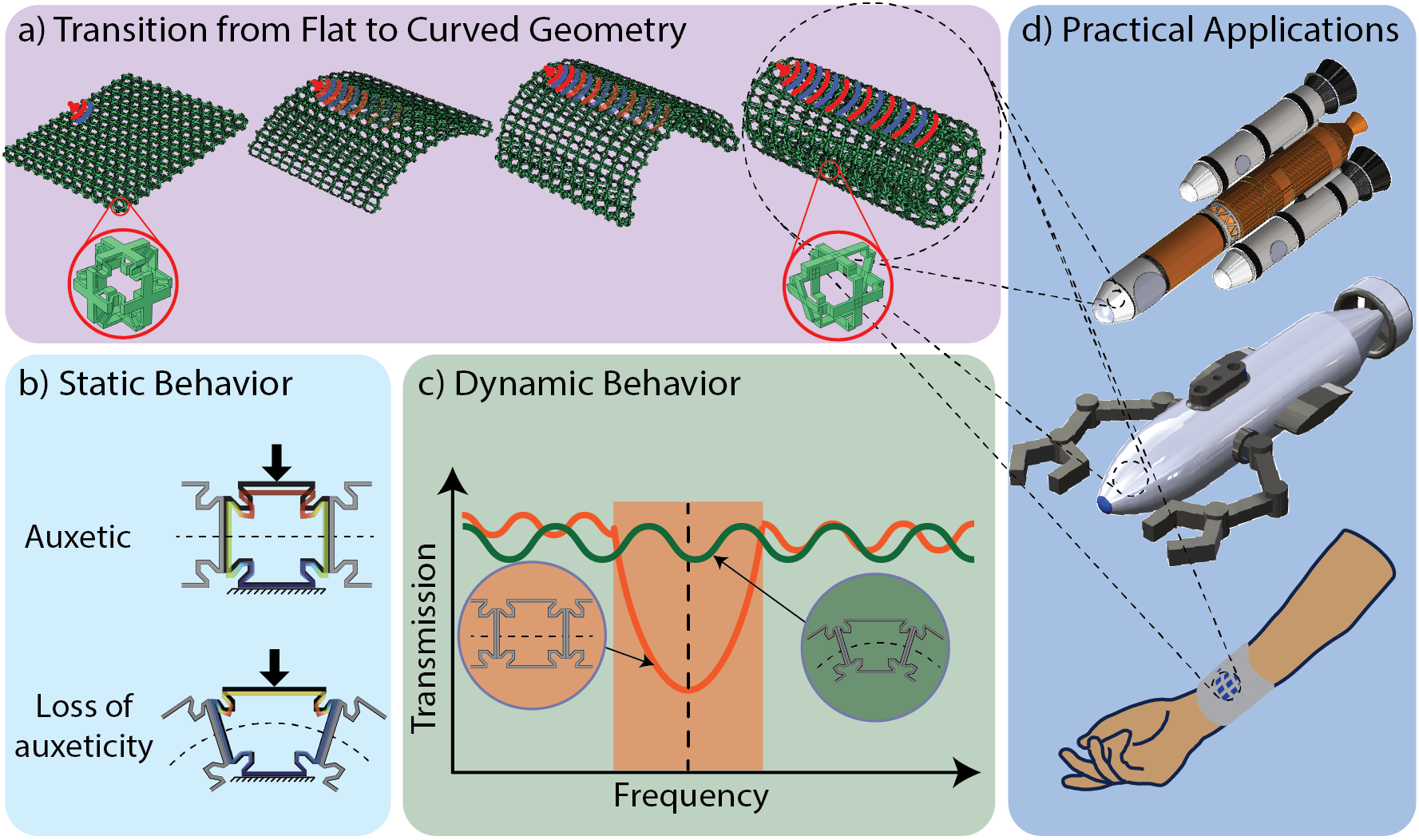}
\textcolor{black}{\caption{\label{fig:Concept}\textbf{Concept:} a) Transition of a structure from a flat plate to fully curved cylinder. b) Effect of curvature on the static behavior of metamaterials, where a flat auxetic structure loses its auxeticity due to curvature. c) Effect of curvature on changing the dynamic response of metamaterials, where a confirmed band gap for a flat geometry might vanish in the presence of curvature. d) Curvature-demanding potential applications of metamaterial.}}
\end{figure*}

Metamaterials are artificially designed structures that consist of a periodic arrangement of basic building blocks (i.e., unit cells) that repeat in space. Metamaterials can be designed for remarkable static, quasi-static and dynamic properties. Such properties can include being ultra-lightweight \cite{schaedler2011ultralight, deshpande2001foam, deshpande2001effective},  having negative effective Poisson's ratio \cite{gatt2013realistic, babaee20133d, greaves2011poisson, saxena2016three}, polar elasticity \cite{bilal2017intrinsically}, negative effective mass, or negative effective stiffness \cite{huang2009negative, zhou2012elastic}. Metamaterials have been suggested in a wide range of applications including wave guiding, thermal and sound waves control, wave focusing, cloaking, and energy harvesting \cite{ kim2012seismic, zhu2014negative, su2016focusing, cummer2016controlling, qi2016acoustic, bilal2017reprogrammable, ma2020acoustic,  chen2016design, bilal2012topologically, bilal2011optimization, li2021transforming, xu2019metamaterials, bilal2021experimental, zhang2020asymmetric, ning2020active,  gao2022acoustic, gonella2009interplay,zhao2022graded}. One of the most fundamental tools in analyzing  metamaterials' behavior lies in their dispersion relation \cite{maldovan2013sound}, which is a correlation between frequency and wavenumber \cite{deymier2013acoustic}. The dispersion relation contains information about metamaterial properties ranging from quasi-static elastic response (at very low frequencies) to its thermal conductivity (at higher frequencies). Dispersion curves are calculated using an assumption of infinite repetition of a single unit cell along the lattice vectors, while taking into account the unit cell symmetry and its irreducible Brillouin zone (IBZ) \cite{brillouin1953wave}. Both the static and dynamic behavior of metamaterials have been heavily investigated in the literature, including the early calculated dispersion curves \cite{kushwaha1993acoustic, liu2000locally}. Most of the literature, however, considers geometries that repeat in Cartesian coordinates. While the static properties of metamaterials have been considered in the presence of curvature \cite{yang20201d, hua2019multistable, ling2021design}, metamaterials' dynamic properties remain largely unexplored in curved coordinates. For example, the dispersion curves for curved unit-cells have been calculated using projected flat irreducible Brillouin zone (IBZ) \cite{yu2021framework, yao2022flexural, yao2023metamaterial, zheng2023emergence}. In addition, wave propagation in finite curved structures \cite{maurin2017bloch,dai2023vibro}, helical wires \cite{treyssede2007numerical}, and  laminated structures with periodicity in the radial direction has been studied \cite{mace2005finite, ghinet2005transmission, manconi2009wave, kitagawa2009bloch, ma2014band}. Moreover, dispersion curves in the axial direction for curved unit cells at different circumferential modes have been calculated \cite{nateghi2016stopband, nateghi2017wave, hakoda2018using, nateghi2019design, zhou2020two, liu2021computational, an2021metamaterial, jin2022design, manushyna2023application, xu2012low}. \textcolor{black}{Within the limited literature of curved metamaterials, comparisons between the static and the dynamic behavior of the same metamaterial in the flat and curved  domains remain scarce}. In this study, we conduct both numerical and experimental analysis to examine the change in the effective static and dynamic properties of metamaterials in the presence of curvature. \textcolor{black}{For the static property, we consider the effective Poisson's ratio of the metamaterial [Fig.\ref{fig:Concept} b], while for the dynamic property, we consider band gap frequency ranges [Fig.\ref{fig:Concept} c]. Both properties, static and dynamic, are important characteristics of metamaterials that we study their sensitivity to curvature}. We present dispersion calculations of curved unit-cells accounting for both axial and circumferential modes. Our analysis could facilitate the integration of metamaterials in applications with curvature-presence, such as aerospace structures, underwater vehicles, automotive components, and wearable devices [Fig. \ref{fig:Concept} d]. We consider unit-cell designs exhibiting different effective Poisson's ratios ranging from negative to zero to positive values \cite{roshdy2023tunable}. We analyze the different Poisson's ratio designs in three scenarios: (1) a flat unit cell, with identical features in all directions, repeated periodically to form a flat structure, (2) quasi-curved unit cell, with distorted (i.e., asymmetrical) unit cell repeated periodically to form a flat structure, and (3) curved unit cell, with the same distorted (i.e., asymmetrical) unit-cell forming a curved structure. We calculate the change in the effective static properties (i.e., Poisson's ratio) and the effective dynamic properties (i.e., wave attenuation capacity) using numerical simulations. \textcolor{black}{We investigate the similarities between the quasi-curved and the curved scenarios in both static and dynamic behavior}. In addition, we fabricate metamaterial structures, both flat and curved, with negative and positive Poisson's ratio. We experimentally characterize the dynamic response of the realized samples through harmonic wave excitation and scanning laser Doppler vibrometry  to verify our findings. 

\begin{figure*}
\centering
\includegraphics[width= \textwidth]{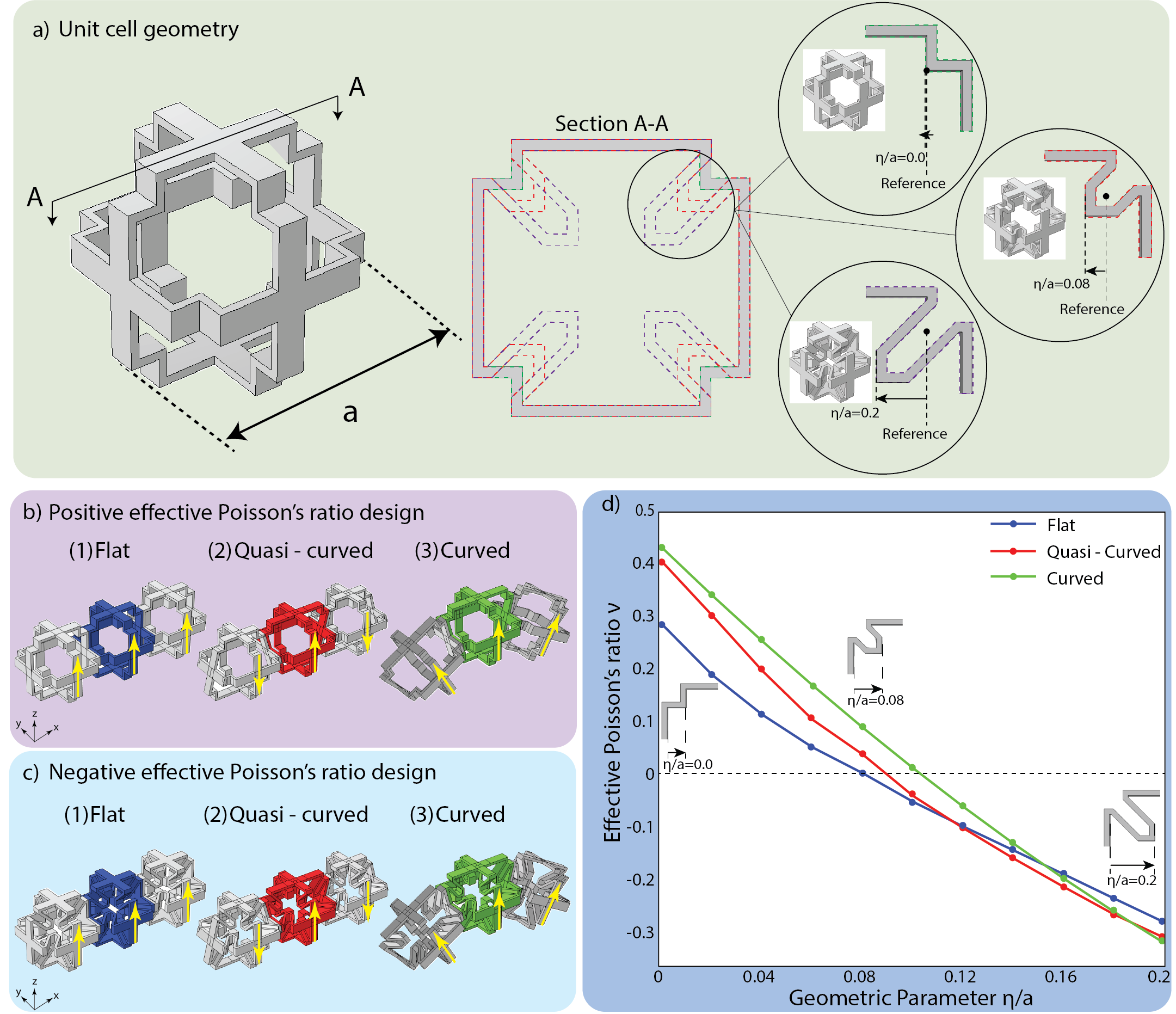}
\caption{\label{fig:Poisson's ratio}\textbf{Poisson's ratio analysis:} \textcolor{black}{a) Schematic of the unit cell geometry with the variation of $\eta /a$.} b) 3 unit cells of flat, quasi-curved and curved geometry with positive effective Poisson's ratio. c) 3 unit cells of flat, quasi-curved and curved geometry with negative effective Poisson's ratio. d) Effective Poisson's ratio as a function of unit cell corner design $\eta /a$, where a is the unit cell lattice constant.}
\end{figure*} 

\section{Static analysis }
We consider three-dimensional unit cells with a geometric feature that allows for the tunability of the effective Poisson's ratio, $\nu$. By adjusting the shape of the unit cell corner, each design can exhibit a different value of $\nu$. We quantify the shape of the corner with a parameter ($\eta/a$), where $a$ is the unit cell lattice constant, \textcolor{black}{and $\eta$ is the horizontal length of the corner feature measured from the reference point where $\eta$ = 0}. for a unit cell with a right-angle corner [Fig. \ref{fig:Poisson's ratio} a]. We analyze the static behavior of each design in the presence of three different levels of curvature: (1) flat unit cell forming a flat structure (i.e., flat), (2) distorted unit cell forming a flat structure (i.e., quasi-curved), and (3) distorted unit cell forming a curved structure (i.e., curved). For case (1), we construct our flat structure by repeating the cubic unit cell in a straight line in the x-direction, forming a beam. Then we repeat the resulting beam in the y-direction creating a plate  with a single unit cell in thickness. For case (3), we construct our curved structure by repeating a distorted version of the cubic unit cell circumferentially such that it forms a closed ring with $N$ number of repetitions. Then we repeat the resulting ring in the axial direction, creating a cylinder that is one unit cell in thickness. For case (2), we utilize the same distorted unit cell (i.e., from the curved structure in case 3), however, we alternate the unit cells top and bottom such that they form a flat beam. Then we repeat the resulting beam in the y-direction creating a plate that is one unit cell in thickness [Fig. \ref{fig:Poisson's ratio} b,c]. We identify one design from each category (+ve of $\nu$, 0 $\nu$, and -ve of $\nu$) for flat unit cells. The unit cell with positive $\nu$ has a value of $\eta/a = 0$, the design of zero $\nu$ has a value of $\eta/a = 0.08$, and finally, the design with negative $\nu$ has a value of $\eta/a = 0.2$.
 
We begin our study with the flat case as a baseline for the other two cases of quasi-curved and curved structures. We study the interplay between curvature and unit cell corner design parameter ($\eta/a$) by calculating the effective Poisson's ratio $\nu$ numerically using the finite element method through COMSOL Multi-physics (version 6.1). In all numerical simulations, we consider the mechanical properties to be: (density $\rho = 1205$ $kg/m^3$, Young's modulus E = 1.22 GPa, and Poisson's ratio $\nu = 0.3$). We apply a compression load to a 10x10x1 flat finite structure in the x-direction, while allowing the y-direction to deform freely. \textcolor{black}{The applied compression load is 3\%a strain magnitude. However, the actual magnitude of the strain has a negligible effect on the value of $\nu$  from 1-4\%a (See supplementary figure S1)}. We vary the value of ($\eta/a$) from 0 to 0.2, with a fixed lattice constant, $a = 12.7$ mm. \textcolor{black}{For values of $\eta/a$ higher than 0.2, the unit cell undergoes either self collision when compressed, or self collision in the design itself \cite{roshdy2023tunable}}. We calculate the average deformation in the y-direction for each design and plot the resulting $\nu$ as a function of  ($\eta/a$) [Fig. \ref{fig:Poisson's ratio} d]. Our reference positive design has a value of $\nu =$ 0.284, where the structure expands in the y-direction due to the compression in the x-direction. Conversely, the flat negative design has a value of $\nu =$ -0.283, resulting in contraction in the y-direction with the compression in the x-direction.

In order to analyze the effect of curvature on the static behavior of the metamaterial, we construct a finite structure with 12 unit cells curved circumferentially to form a ring with an inner radius of 38.1 mm and an outer radius of 63.5 mm. The ring structure is then repeated 10 times in the axial direction to create a cylinder. To calculate the effective Poisson's ratio, we apply a compression load to the structure in the axial direction, while allowing the cylinder's circumferential surface to deform. We plot $\nu$ as a function of ($\eta/a$) for both the curved and the quasi-curved cases. We observe a clear change in the effective $\nu$ for the same ($\eta/a$). For the curved positive $\nu$ design, the effective Poisson's ratio increases by 62$\%$ ($\nu = 0.459$) causing the cylinder diameter to expand under an axial compression load. While the curved negative $\nu$ design exhibits 11 $\%$ ($\nu = -0.315$) change in the effective Poisson's ratio, leading to a contraction of the cylinder diameter under axial compression. The value of effective Poisson's ratio for the zero $\nu$ design changes from $\nu = 0$ in the flat case to $\nu = 0.113$ in the case of curved structure. Finally, the quasi-curved geometries exhibit a static behavior that deviates from both the flat and curved cases. The positive $\nu$ design has 42 $\%$ ($\nu = 0.405$) increase in the effective Poisson's ratio, while the negative $\nu$ design has similar behavior to the curved case with about 11 $\%$ change ($\nu = -0.315$). The zero $\nu$ design has a value of effective Poisson's ratio that changes from 0 in the flat case to ($\nu = 0.03$) in the quasi-curved case. [Fig. \ref{fig:Poisson's ratio} c]. The conducted static analysis shows that effective Poisson's ratio of the considered metamaterials changes significantly for the same design in the presence of curvature. The quasi-curved structures show different behavior compared to both flat and curved cases, however, its behavior is closer to the curved structures, particularly for higher values of ($\eta /a$).

\section{Dynamic analysis }
To capture the influence of curvature on the dynamical properties of our metamaterials, we calculate the unit cell dispersion curves for both flat and curved cases. Dispersion curves represent the relationship between frequency $\omega$ and wavenumber $\kappa$, considering one single unit cell repeated periodically in space. We apply Bloch's boundary condition to represent the periodic arrangement of our unit cells. 

\subsection{Flat unit cell analysis}
\begin{figure*}
\centering
\includegraphics[width= \textwidth]{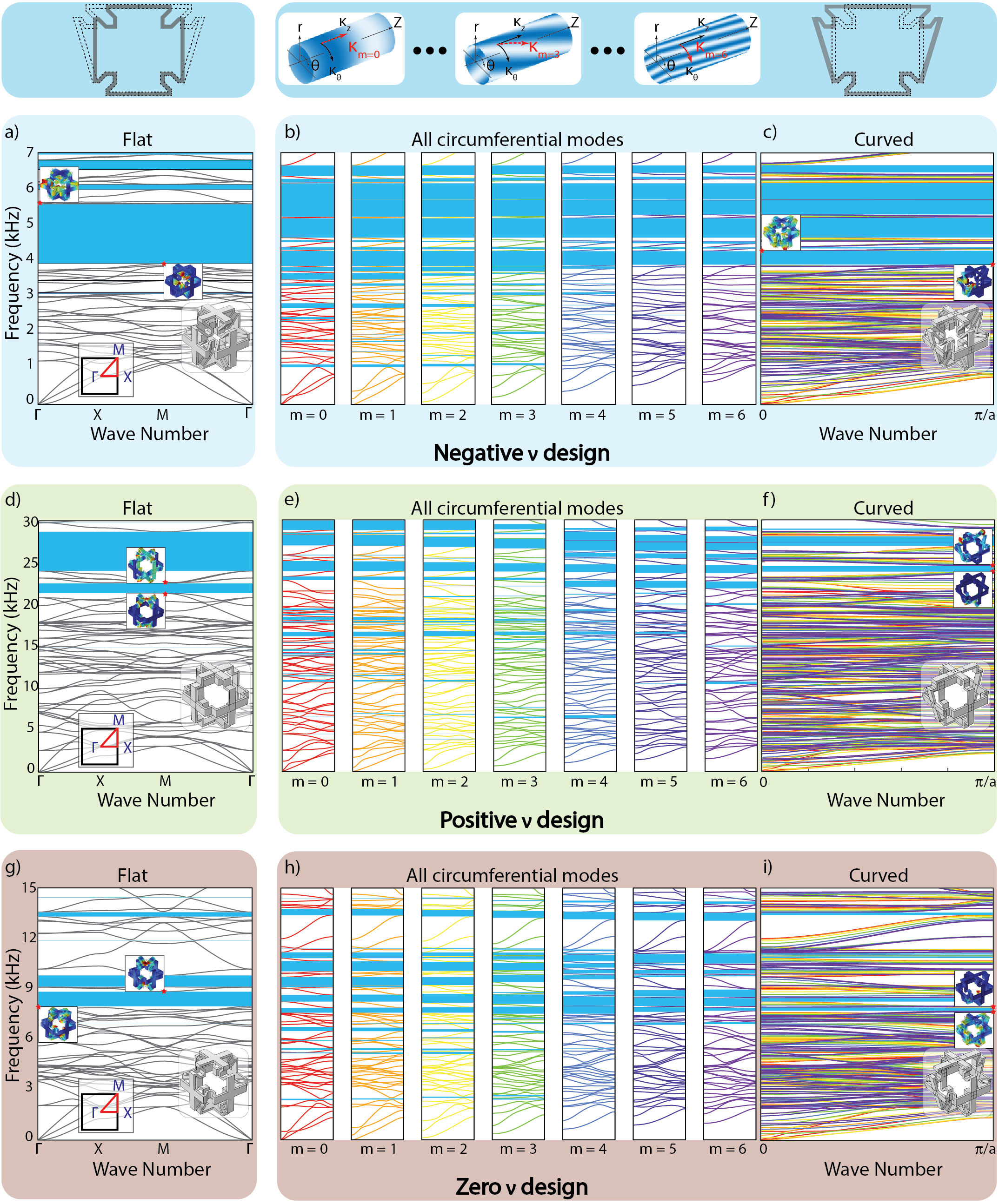}
\caption{\label{fig:Dispersion}\textbf{Dispersion Relations:} a) Dispersion curves for negative $\nu$ flat unit cell. b) Dispersion curves for negative $\nu$ curved unit cell at all circumferential modes. C) Dispersion curves for negative $\nu$ curved unit cell. d) Dispersion curves for positive $\nu$ flat unit cell. e) Dispersion curves for positive $\nu$ curved unit cell at all circumferential modes. f) Dispersion curves for positive $\nu$ curved unit cell. g) Dispersion curves for zero $\nu$ flat unit cell. h) Dispersion curves for zero $\nu$ curved unit cell at all circumferential modes. i) Dispersion curves for zero $\nu$ curved unit cell.}
\end{figure*}

To calculate the dispersion curves for the flat unit cell designs in Cartesian coordinates (x,y,z), we solve the resulting eigenvalue problem $(-\omega^2 M + K(\kappa))\bar{U} = 0$, while assuming a solution in the form: \textcolor{black}{$u_{car}(X,\kappa;t) = \bar{u}_{car}(X,\kappa) e^{i n(\kappa_x a + \kappa_y a)} e^{i\omega t}$}, where $M$ is the mass matrix, $K$ is the stiffness matrix, \textcolor{black}{$\bar{U}$ is the eigenvector}, $u_{car}$ is the displacement field in the Cartesian coordinates, \textcolor{black}{$\bar{u}_{car}$ is the displacement Bloch function in Cartesian coordinates, $X$ is the position vector in Cartesian coordinates}, $\kappa_x$ and $\kappa_y$ are the wavenumbers in the $x-$ and $y-$direction, respectively, $n$ is the unit cell index, $a$ is the lattice constant, $t$ is time, $\omega$ is the angular frequency, and $i= \sqrt{-1}$ is the imaginary number. We account for periodicity in both $x-$ and $y-$directions. Due to the symmetry of the flat unit cell design, we consider the irreducible Brillouin zone (IBZ) ($\Gamma-X-M-\Gamma$). We numerically calculate the dispersion curves using the finite element method utilizing COMSOL Multiphysics. For case (1), i.e., flat, we consider the dispersion curves of three designs with negative, positive, and zero $\nu$. For the unit cell with a negative $\nu$ value, we observe three distinct band gap regions in the frequency range between 0 and 7 kHz. These frequency ranges are where waves cannot propagate through the structure because there is no real solution to the eigenvalue problem [Fig. \ref{fig:Dispersion}a].  As ($\eta/a$) decreases, both the effective unit cell stiffness and Poisson's ratio increase, resulting in an overall increase of the band gap frequency ranges. For the unit cell with zero $\nu$, there are three prominent band gaps within the frequency range of 0-15 kHz [Fig. \ref{fig:Dispersion}g]. For the unit cell with the positive $\nu$, we observe two distinct band gaps within the frequency range 0-30 kHz [Fig. \ref{fig:Dispersion}d]. In addition, we plot the mode shapes of the three flat designs at the upper and lower boundaries of the first band gap [Fig. \ref{fig:Dispersion}]. To capture the influence of the stiffness change at different $\nu$ values on frequency, we consider the  frequency range of the lowest dispersion branch for each design. For the design with negative $\nu$, the lowest dispersion branch peaks at 1.1 KHz. For for unit cell with zero $\nu$, this value increases to 2 kHz. While for the unit cell with positive $\nu$, it is 3.1 kHz. The increase in frequency is primarily due to the increase in effective stiffness as ($\eta$) decreases.

\subsection{Curved unit cell analysis}
In order to calculate the dispersion curves for metamaterials with curved unit cell designs in cylindrical coordinates (r,$\theta$,z), we assume periodicity in the axial direction, we consider a helical wave propagation, assuming the Bloch form of wave solution \textcolor{black}{$u_{cyl}(\bar{R},\kappa;t) = \bar{u}_{cyl}(\bar{R},\kappa) e^{i\kappa_z na} e^{i m\theta} e^{i\omega t}$}, where \textcolor{black}{$u_{cyl}$ is the displacement field in cylindrical coordinates, $\bar{u}_{cyl}(\bar{R},\kappa)$ is the displacement Bloch function in cylindrical coordinates, $\bar{R}$ is the position vector in cylindrical coordinates}, $\kappa_z$ is the wavenumber in the axial direction, $\theta$ is the cylindrical coordinate angle, and $m$ is an integer number corresponding to the circumferential mode (Azimuthal mode number). The value of $m$ changes from $0$ to $N/2$ \textcolor{black}{with an increment of 1}, where $N$ is the total number of unit cells around the circumference. We analyze the dispersion for the curved unit cells in the axial direction by sweeping the axial wavenumber $\kappa_z$ from 0 to $\pi /a$, for all values of the circumferential mode number, $m$, ranging from 0 to $N/2$.

\begin{figure*}
\centering
\includegraphics[width= \textwidth]{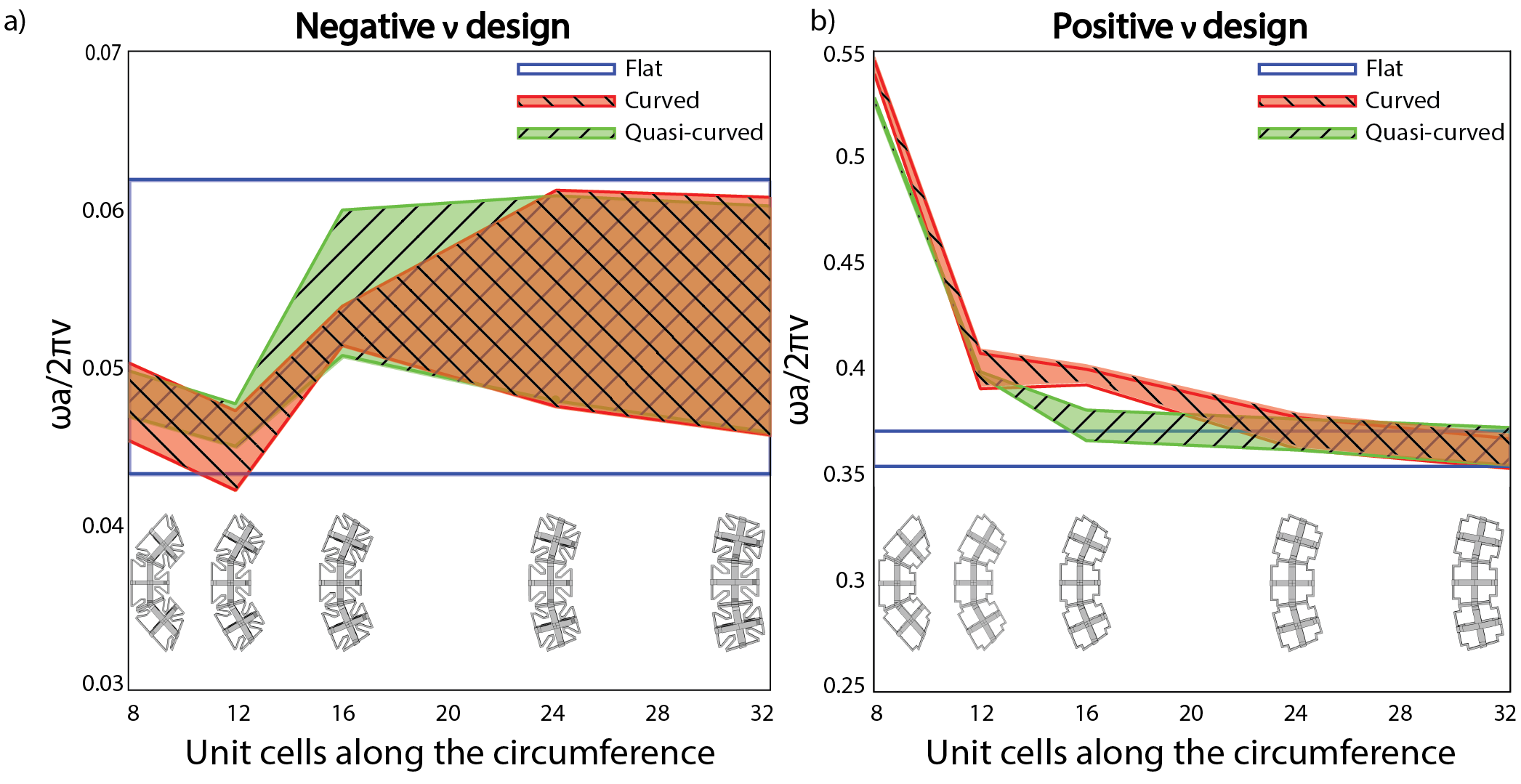}
\caption{\label{fig:N_unit_cells}\textbf{Effect of number of circumferential unit cells:} Change in the frequency of the upper and lower boundaries of the first band gap for curved unit cell (red shaded region), and quasi-curved unit cell (green shaded region) with the change in the number of circumferential unit cells from 8 to 
32, compared to the first band gap of the flat unit cell (blue lines) for a) negative $\nu$ design, and b) positive $\nu$ design.}
\end{figure*}

For the fully curved case, i.e., case (3), we consider the dispersion curves of three different curved designs with negative, positive, and zero $\nu$. For the unit cell with negative $\nu$, we identify five band gap regions in the frequency range 0-7 kHz. The dispersion curves for the same design in flat coordinates exhibits three band gaps at different frequency ranges. In addition, we plot the unit cell vibrational modes at the boundaries of the first band gap of the curved unit cell design. The mode shapes show different displacement profiles compared to the corresponding flat unit cell vibrational modes [Fig. \ref{fig:Dispersion}c]. In order to distinguish the contribution of helicity to the dispersion curves of the curved unit cell, we plot the dispersion curves of each circumferential mode separately (i.e. $m = 0$ to $6$). The dispersion curves for $m=0$ represent no helicity, translating to waves only propagating in the axial direction. The dispersion curves for $m=6$ represent waves propagating with maximum helicity (See Video 1). For intermediate values of $m$, the helical waves propagate with varying angles of helicity [Fig. \ref{fig:Dispersion}b]. For the curved unit cell design with positive $\nu$, we observe a clear change compared to the flat design, also with positive $\nu$. The width of the band gaps shrinks significantly, showing multiple narrow band gaps. In addition, the mode shapes at the boundaries of the first band gap reveal a noticeable difference compared to the flat dispersion curves [Fig. \ref{fig:Dispersion}f]. Moreover, we plot the separate dispersion curves for the curved unit cell with the positive $\nu$ design for different circumferential modes. We observe more band gaps with wider frequency ranges at $m=0$, however, the width of the band gaps decreases gradually as the circumferential mode index $m$ increases [Fig. \ref{fig:Dispersion}e]. In addition to the curved unit cell designs with both negative and positive $\nu$, we calculate the dispersion curves for a curved unit cell design with zero $\nu$. We observe a significant  change in the dispersion curves from the design with flat unit cell compared to the design with curved unit cell, with a significant decrease in the width of the band gaps [Fig. \ref{fig:Dispersion}i]. Furthermore, we plot the separate dispersion curves for each circumferential mode for the zero $\nu$ design. We observe a reduction in the number and  width of the band gaps' frequency ranges with the increase of the circumferential mode index from $m=0$ to $m=6$ [Fig. \ref{fig:Dispersion}h].

\subsection{Quasi-curved unit cell analysis}
\begin{figure*}
\centering
\includegraphics[width= \textwidth]{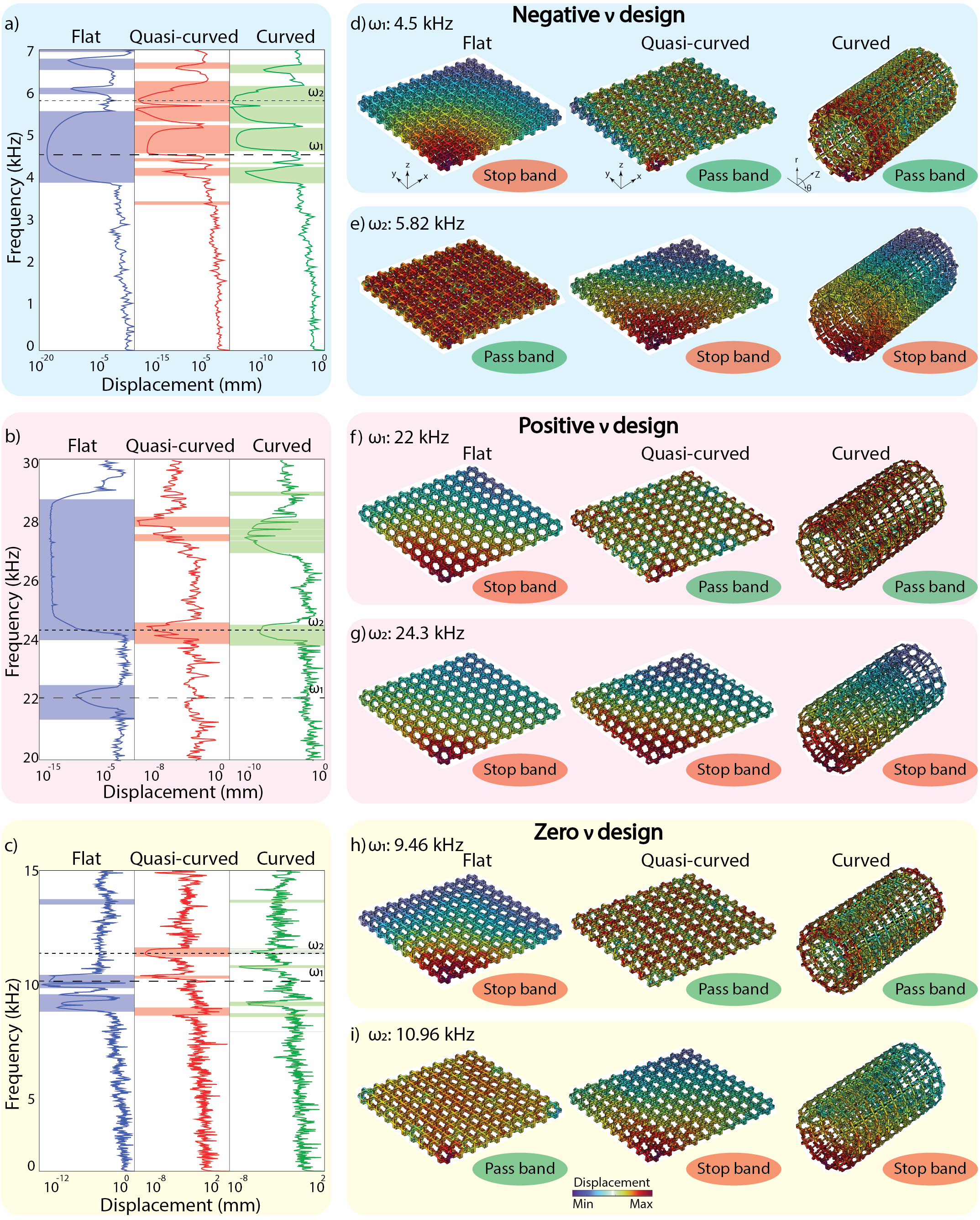}
\caption{\label{fig:Numerical - FRF}\textbf{Numerical FRF:} Finite structure frequency response function for flat, quasi-curved, and curved structures of a) negative $\nu$ design, b) positive $\nu$ design, and c) zero $\nu$ design. d-i) Finite structures mode shapes at different excitation frequencies.}
\end{figure*}

To calculate the dispersion curves of the quasi-curved designs, we consider a super cell of 2x2 quasi-curved unit cells, where 2 distorted unit cells are repeated in an alternating fashion in the $x$-direction, and then we repeat these 2 unit cells in the $y$-direction without alternation. Such construction allows us to use the flat dispersion calculation method for the distorted  unit cells without having a curved structure. We solve the eigenvalue problem considering the Bloch form of the solution with periodic boundary conditions in both the $x$- and $y$-direction. We consider the irreducible Brillouin zone ($\Gamma-X-M-\Gamma$). We calculate the dispersion curves for three designs with varying Poisson's ratios and record the regions of the band gaps. \textcolor{black}{We plot the band gap regions for the quasi-curved as shaded regions over the finite structure numerical frequency response functions} [Fig. \ref{fig:Numerical - FRF} a, b, and c].\\

\subsection{Effect of circumferential number of unit cells on dispersion curves}
In order to investigate the effect of the number of unit cells in the circumferential direction (i.e., curved design with different curvature angles) on the metamaterials' dynamics, we vary the number of unit cells arranged circumferentially from 8 to 32 for both the negative and the positive $\nu$ designs. In each case, we calculate the dispersion curves for the curved unit cell design in the axial direction at all circumferential modes. In addition, we calculate the dispersion curves for the corresponding flat unit cell design, and the quasi-curved unit cell design. We plot the first band gap for flat, quasi-curved, and fully curved unit cells in normalized frequency to account for the change in scale. We define the normalized frequency as  ($\omega a/2\pi v$), where $v$ is the longitudinal speed of sound in the material. We measure the difference in band gap frequency ranges between curved and quasi-curved cases versus the flat case. The mismatch is calculated for curved design as $|\omega_{curved} - \omega_{flat}|/\omega_{flat} \%$ and for quasi-curved design as $|\omega_{quasi-curved} - \omega_{flat}|/\omega_{flat} \%$. In the case of negative $\nu$ design with curved unit cells, we observe a decrease in the mismatch for the upper boundary of the band gap from $18.4\%$ for 8 unit cells to $1.8\%$ for 32 unit cells. Additionally, the mismatch of the lower boundary of the band gap remains around $5\%$. In the case of the negative $\nu$ design with quasi-curved geometry, the mismatch decreases from $19\%$ to $2.7\%$ for the upper boundary of the band gap, while for the lower boundary, the mismatch decreases from $8.3\%$ to $5.9\%$ [Fig. \ref{fig:N_unit_cells}a]. In the case of the positive $\nu$ design, the mismatch of the curved band gap upper boundary decreases from $47\%$ for 8 unit cells, to less than $1\%$ for 32 unit cells, and the mismatch of the lower boundary decreases from $52\%$ to $0.3\%$. For the positive $\nu$ design with quasi-curved geometry, the mismatch of the upper boundary decreases from $42\%$ to $0.4\%$, while for lower boundary, the mismatch decreases from $48.1\%$ to $0.17\%$ [Fig. \ref{fig:N_unit_cells}b]. In other words, as the number of unit cells in the circumferential direction increases, the band gap frequency for the curved (and quasi-curved) designs matches that of the flat design. The curvature of the unit cell has a significant effect on the metamaterials dynamics particularly for lower number of unit cells in the circumferential direction. Furthermore, the effective dynamical behavior of the quasi-curved unit cells is similar to that of the curved unit cells, despite being flat structure with simpler dispersion curve calculation compared to the curved case. Such similarity could simplify the analysis of curvature significantly, as it is done in Cartesian coordinates, while capturing the dynamics of the curved metamaterials.

\subsection{Finite structure numerical analysis}

In order to validate our infinite unit cell model, we perform numerical simulations for \textit{finite} structures. We calculate the frequency response functions, i.e., FRFs, for the three designs considered earlier in the manuscript with negative, positive, and zero effective Poisson's ratios, $\nu$. For each design, we evaluate the wave propagation characteristics of the three cases: (1) flat, (2) quasi-curved, and (3) curved. For the negative $\nu$ design, in the flat configuration, i.e., case (1), we consider a plate composed of 10x10x1 unit cells with unit cell lattice constant $a = 12.7 mm$. We excite the plate from one corner of the metamaterial with an elastic wave, and we measure the transmitted wave at the diagonally opposite corner of the metamaterial. For case (3), i.e., curved,  we repeat 12 unit cells circumferentially forming a cylindrical ring with a thickness of one unit cell.  Then, we repeat the ring 10 times in the axial direction, forming a cylinder. We excite an elastic wave at the lower unit cell in the first layer and measure the response at the upper unit cell on the opposite side of the cylinder. In addition to the flat and the curved configurations, we consider a quasi-curved finite metamaterial, i.e., case (2), where the curved unit cell is repeated in a flat manner rather than along a curved ring. To construct our quasi-curved metamaterial, we repeat the distorted unit cell in one direction (x-direction) considering alternating orientation in the z-direction (out of plane direction), creating a quasi-curved beam. Then, we repeat the quasi-curved beam 10 times in the y-direction. We excite the quasi-curved metamaterial at one corner, and measure the transmitted wave at the opposite corner diagonally. For all considered structures, the excitation source is a chirp signal that sweeps over the calculated frequency range in the infinite model. We plot the FRFs in a semi-log scale for the displacements as a function of excitation frequency [Fig. \ref{fig:Numerical - FRF}a]. The FRFs reveal frequency ranges with high and low wave transmission amplitudes. We also calculate the FRFs for both the positive [Fig. \ref{fig:Numerical - FRF}b] and zero [Fig. \ref{fig:Numerical - FRF}c] $\nu$ design in all three cases: flat, quasi-curved, and curved. For all three simulated designs with negative, positive and zero effective $\nu$, both wave transmission and attenuation ranges are in a good agreement with the predicted band gap frequency ranges (i.e., shaded regions) using the infinite models. The dispersion curves for the (1) flat, (2) quasi-curved, and (3) curved instances are confirmed by matching attenuation regions in the FRFs and the unit cell band gap ranges. Additionally, it is worth noting that the FRF and band gap frequency ranges for the quasi-curved geometry are closer to the band gap frequency ranges of the curved geometry rather than the flat ones. These results indicate that the curvature effect is primarily geometric and the quasi-curved model, in Cartesian coordinates, can be used to predict some of the dynamical characteristics of curved metamaterial designs  with moderate levels of curvature.

To further demonstrate the effect of curvature and help in visualizing its significance on the effective dynamical properties of our metamaterials, we plot the mode shapes of the three considered structures for flat, quasi-curved and curved designs for each of the three considered $\nu$ values. For the negative $\nu$ design we consider two frequencies: 4.5 kHz, which is a stop band for the flat structure, but a pass band for  both the quasi-curved and the curved structures [Fig. \ref{fig:Numerical - FRF}d]. The second frequency is 5.82 kHz, which is a pass band frequency for the flat structure, yet a stop band for both the quasi-curved and the curved structures [Fig. \ref{fig:Numerical - FRF}e]. For the positive $\nu$ design we consider two frequencies: the first is 22 kHz, which is a stop band frequency for flat geometry, but it is a pass band for both the quasi-curved and curved structures [Fig. \ref{fig:Numerical - FRF}f]. The second frequency is 24.3 kHz, which is a stop band frequency for the flat, quasi-curved, and curved structures [Fig. \ref{fig:Numerical - FRF}g]. Lastly, for the zero $\nu$ design, we consider the two excitation frequencies: 9.46 kHz which is a stop band for the flat structure, switching to a pass band for the quasi-curved and curved structures [Fig. \ref{fig:Numerical - FRF}h]. The second frequency is 10.96 kHz, which is a pass band in the case of the flat structure, yet a stop band in both the quasi-curved and curved structures [Fig. \ref{fig:Numerical - FRF}i]. In all of the demonstrated cases, except for panel g, the structure functionality is reversed in the presence of curvature from transmission to attenuation or vice-versa, which underscores the significance of incorporating curvature in metamaterials' design and analysis.

\section{Experimental realization}
\begin{figure*}
\centering
\includegraphics[width= \textwidth]{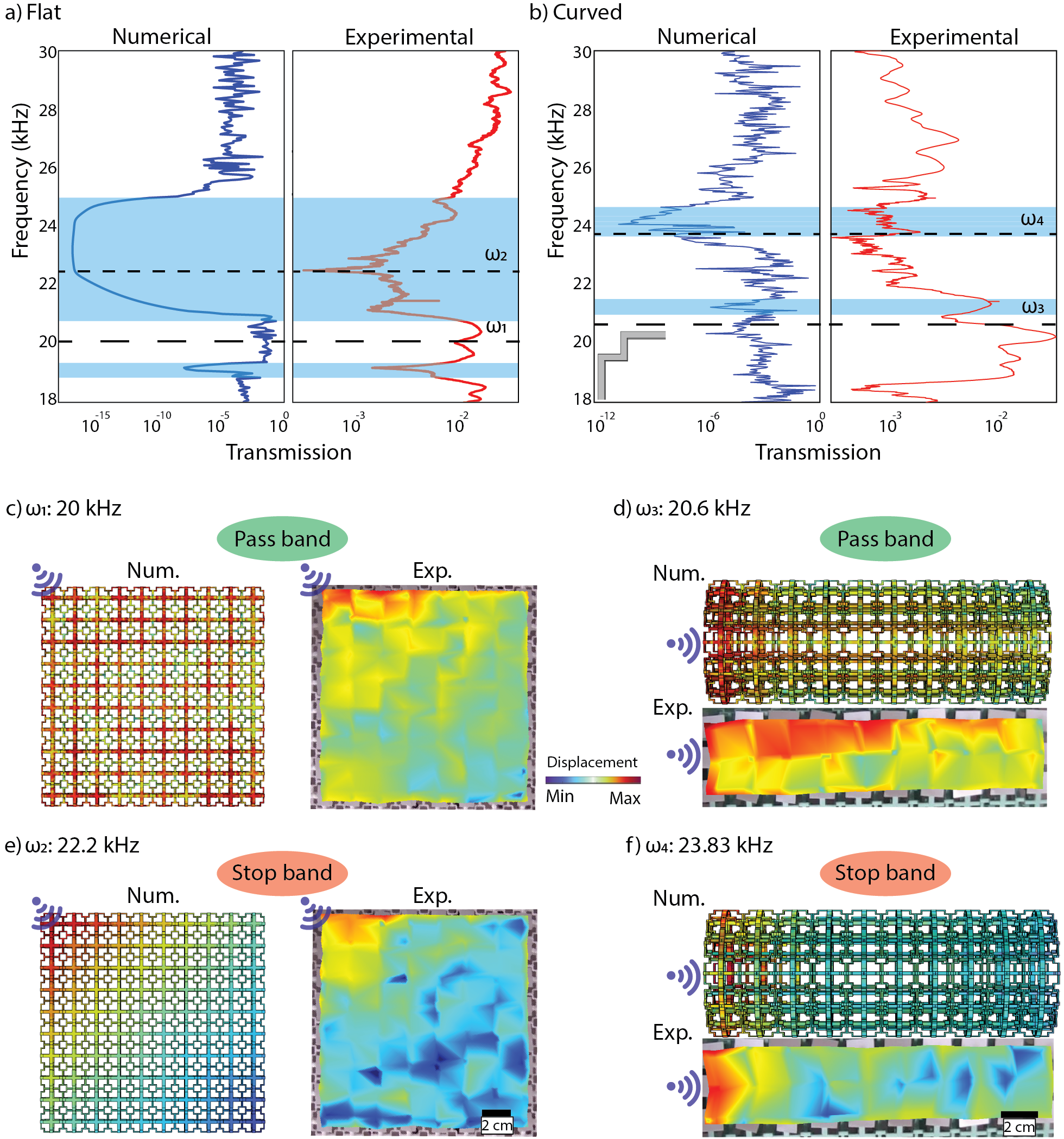}
\caption{\label{fig:Positive - Experiment}\textbf{Positive Poisson's ratio experiment:} a) Flat, and b) curved numerical and experimental FRFs for the positive $\nu$ design with the band gap regions are shaded in light blue. Numerical and experimental mode shapes at excitation frequency of c) 20 kHz for flat structure, and d) 20.6 kHz for curved structure. Numerical and experimental mode shapes at excitation frequency of e) 22.2 kHz for flat structure, and f) 23.82 kHz for curved structure.}
\end{figure*} %[width = \columnwidth]]

\begin{figure*}
\centering
\includegraphics[width= \textwidth]{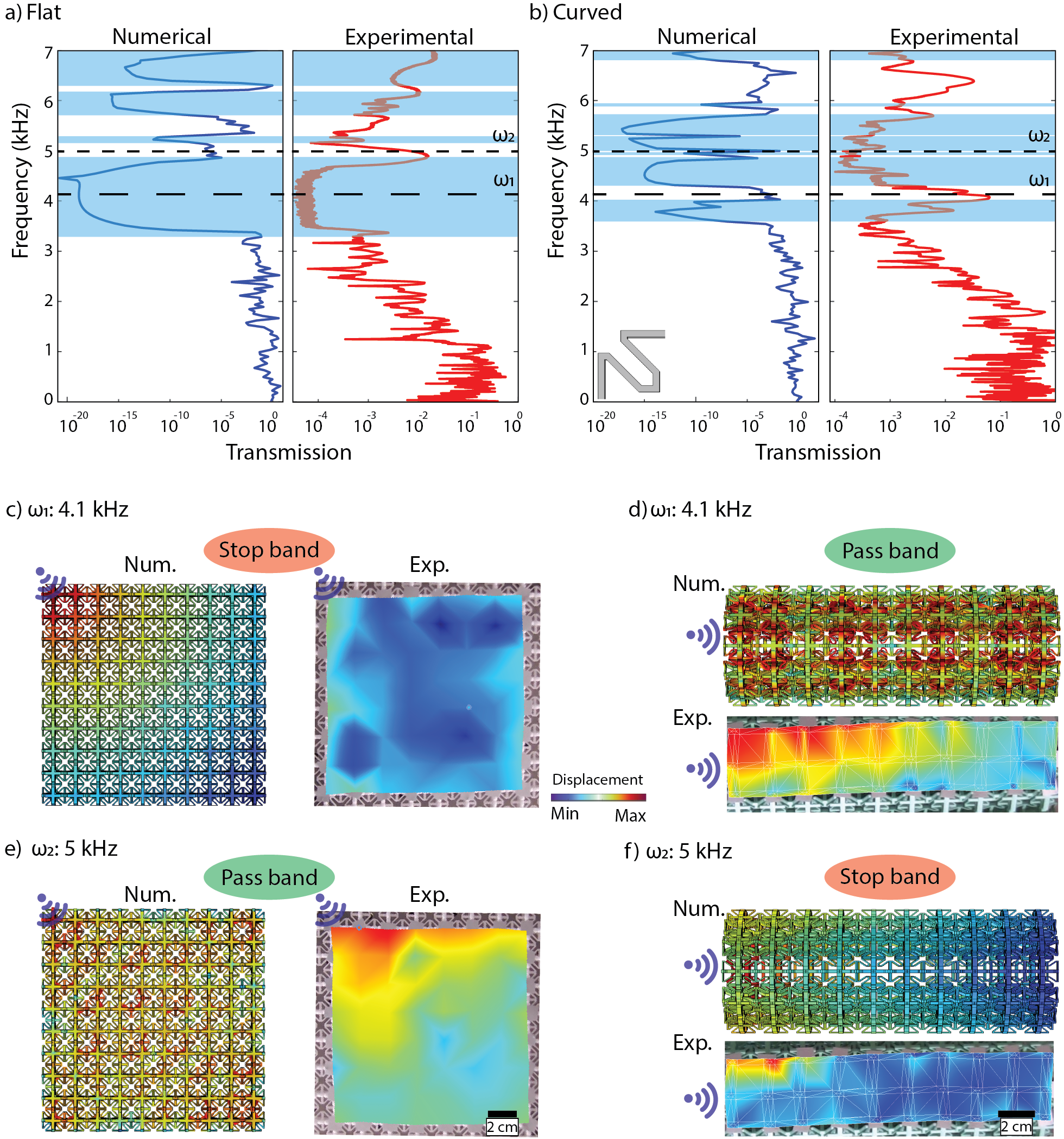}
\caption{\label{fig:Negative - Experiment}\textbf{Negative Poisson's ratio experiment:} a) Flat, and b) curved numerical and experimental FRFs for the negative $\nu$ design with the band gap regions are shaded in light blue. Numerical and experimental mode shapes at excitation frequency of 4.1 kHz for c) flat, and d) curved geometry. Mode shapes at excitation frequency of 5 kHz for e) flat, and f) curved geometry.}
\end{figure*} 

In order to validate our numerical analysis, both infinite and finite, we fabricate multiple prototypes for flat and curved structures constructed from both the positive [Fig. \ref{fig:Positive - Experiment}] and negative [Fig. \ref{fig:Negative - Experiment}]  effective $\nu$ designs. The samples are fabricated using PA12 plastic and realized through selective laser sintering (SLS) additive manufacturing. The mechanical properties of PA12 are: (density $\rho = 930$ $kg/m^3$, Young's modulus E = 1.65 GPa, and Poisson's ratio $\nu = 0.37$), with a fabricated unit cell size of 19 mm. The flat structures are composed of 10x10 unit cells repeated in both x and y-directions with a single unit cell along their thickness. The curved structures are composed of 10 rings repeated axially, where each ring has 12 unit cells along its circumference. We excite the samples with a chirp signal using piezoelectric bending discs (T216-A4NO-05). For the flat samples, we attach the piezoelectric disk at one corner and  measure the displacement at the diagonally opposing corner. For the curved samples, we excite the cylinder at the lower unit cell of the first ring and measure the displacement at the upper unit cell of the opposite side of the cylinder. We utilize a scanning laser Doppler vibrometer (Polytech 500-PSV) to capture the displacement of our samples. We plot the transmission of the elastic wave as a function of the excitation frequency. We define the transmission as the measured output displacement divided by the input excitation amplitude. We calculate numerically the dispersion curves and frequency response functions for both flat and curved cases, considering designs with negative and positive $\nu$ with unit cell size of 19 mm, and material properties of PA12, and we compare these numerical results with the experimental measurements.

To experimentally validate the numerical results of the positive $\nu$ design, the excitation sweep covers a range of frequency from 18 to 30 kHz. The findings show that the infinite unit cell dispersion band gaps (light blue shaded regions) and the experimental FRFs agree well [Fig. \ref{fig:Positive - Experiment}a-b]. Additionally, the experimental and numerical FRFs for both flat [Fig. \ref{fig:Positive - Experiment}a] and curved [Fig. \ref{fig:Positive - Experiment}b]  geometry show good agreement. In order to further verify the single point measurements of our FRFs, we create a mesh grid of points over the flat structure, with 8 points per unit cell, totalling in 800 points for the entire structure. The scanning laser vibrometer measures the displacement at each point of the grid while the structure is being excited. Additionally, we construct a 240 point mesh grid spanning three rows of unit cells in the curved construction. We plot the experimentally scanned mode shapes for the flat structure at an excitation frequency of 20 kHz which is a pass band frequency in the case of the flat plate [Fig. \ref{fig:Positive - Experiment}c], and at 20.6 kHz for the curved structure [Fig. \ref{fig:Positive - Experiment}d]. We also plot the modes shapes at two stop band frequencies, 22.2 kHz for the flat structure [Fig. \ref{fig:Positive - Experiment}e] and 23.83 kHz for the curved structure [Fig. \ref{fig:Positive - Experiment}f]. The mode shapes in both cases show very good agreement between numerical simulations and experiments for both propagation and attenuation cases.

In the case of the unit cell design with negative $\nu$, we sweep the excitation frequency from 1 Hz to 7 kHz [Fig. \ref{fig:Negative - Experiment}]. The experimental results for the flat negative $\nu$ design align  well with the band gap regions predicted by the infinite unit cell dispersion analysis. Additionally, the experimental FRF shows  good agreement with the numerical FRF [Fig. \ref{fig:Negative - Experiment}a-b]. More importantly, we experimentally observe noticeable differences between the measured transmission of the wave in the flat and the curved structures [Fig. \ref{fig:Negative - Experiment} a-b]. To further validate our findings, we scan the mode shapes for the flat and the curved structures at 4.1 kHz, which is a stop band in the case of the flat structure [Fig. \ref{fig:Negative - Experiment}c] and a pass band frequency in the case of the curved structure [Fig. \ref{fig:Negative - Experiment}d]. We also scan the structural mode shape at 5 kHz, which is a pass band frequency in the case of the flat structure [Fig. \ref{fig:Negative - Experiment}e] and a stop band frequency for the curved structure [Fig. \ref{fig:Negative - Experiment}f]. The experimental mode shapes validate the displacement field of the numerical mode shapes. In addition, the mode shapes show the metamaterial's loss of functionality (i.e., instead of attenuation in the flat metamaterial we observe wave propagation in the curved metamaterial) at the frequency of 4.1 kHz due to curvature [Fig. \ref{fig:Negative - Experiment}c-d]. A similar loss of functionality (i.e., instead of wave propagation in the flat metamaterial we measure attenuation in the curved metamaterial) at 5 kHz [Fig. \ref{fig:Negative - Experiment}e-f]. Both scenarios, verified experimentally, underscore the significance of curvature on the functionality of metamaterials. 

\section{Conclusion}

Our investigation of  the effective static and dynamic properties of metamaterials in the presence of curvature can inform future research in the design of functional metamaterials. In the statics domain: we analyze different unit cells with values of effective Poisson's ratio $\nu$, varying from negative to zero to positive. We conduct our analysis considering three levels of curvature transitioning from flat to quasi-curved to curved structures. We demonstrate the change in the effective Poisson's ratio with the unit cell geometric design parameters at  different curvature levels. In the dynamics domain: we predict the dynamics of flat and curved unit cells as both  infinite and finite structures. Furthermore, we study the effect of the number of unit cells on the dynamical characteristics of the curved and quasi-curved structures compared to the flat structures. We validate our infinite single unit cell dispersion curves against the finite structures' frequency response functions both numerically and experimentally. Our findings demonstrate the importance of accounting for curvature in the analysis of the static and dynamic properties of metamaterials and could provide the means to consider such important factors in the future design of practical metamaterials for functional use.

 \newpage
\beginsupplement
\begin{widetext}
\newpage\hspace{-3mm}\Large{\textbf{Supporting Information: \\}}
\Large{\textbf{Static and dynamic analysis of auxetic three-dimensional curved\\ metamaterials in both axial and circumferential directions}\\}
\\

\textcolor{black}{In order to study the effect of curvature on the static behavior of our metamaterial, we calculate the effective Poisson's ratio of a finite structure of metamaterial. For the flat unit cell design, we construct a plate of 10x10x1 (in x-y-z directions, respectively) unit cells. We apply a fixed boundary condition to the left side of the x-direction of the the finite structure, and apply a prescribed strain (i.e., displacement) to the right side of the x-direction. We calculate the deformation of the y-direction sides (i.e., the lateral deformation). The ratio between the lateral strain in the y-direction to the applied strain in the x-direction is considered as the effective Poisson's ratio of the lattice.} 

\begin{figure}[h]
\centering
\includegraphics{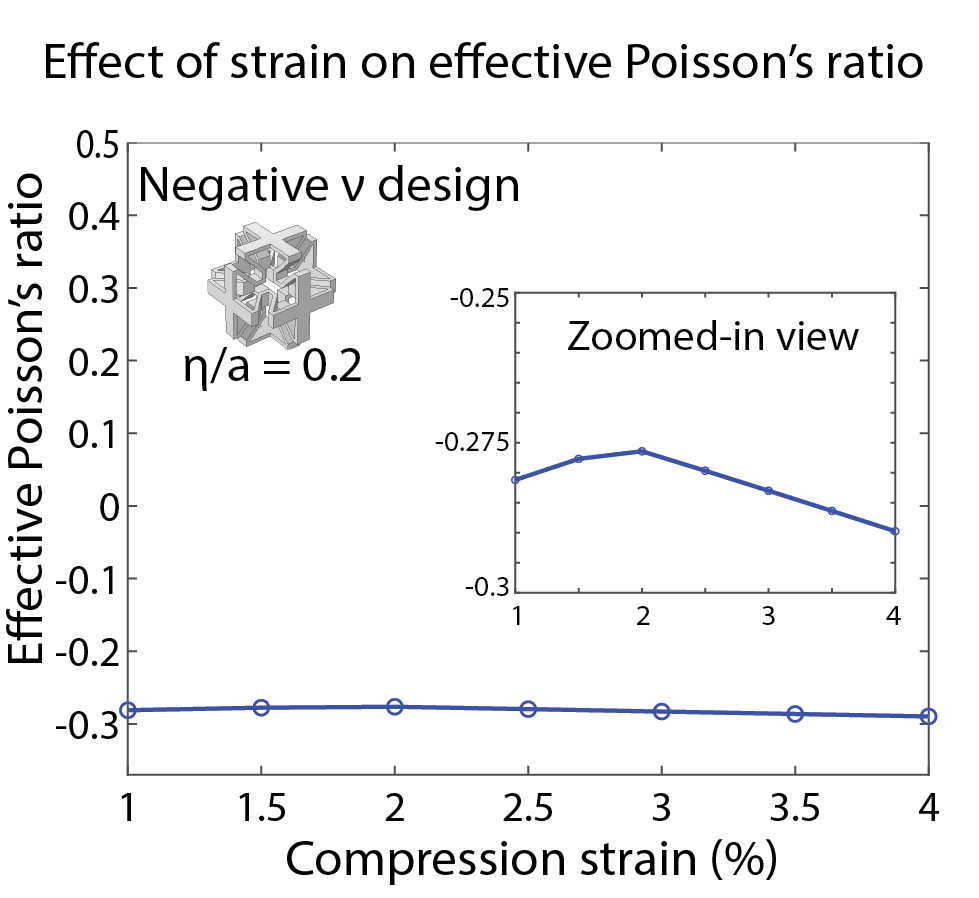}
\caption{\label{fig:S1} Effect of strain's magnitude on effective Poisson's ratio.} 
\end{figure}

\textcolor{black}{In order to study the effect of various applied strain magnitudes on the resultant Poisson's ratio, we consider a finite structure of a unit cell design with $\eta / a = 0.2$. We vary the applied prescribed displacement  from 1 to 4 $\%$a. We calculate the effective Poisson's ratio as a function of the applied strain. We observe a negligible effect  on the effective Poisson's ratio as a function of the applied strain [Fig. \ref{fig:S1}].
In order to obtain the local effective Poisson's ratio at each unit cell in the side free to deform (y-direction side), we plot the local effective Poisson's ratio at each unit cell for the three design with $\eta /a$ = 0.0, 0.08, and 0.2. We consider the average value in dashed lines for each design [Fig. \ref{fig:S2}].} 

\begin{figure}[h]
\centering
\includegraphics{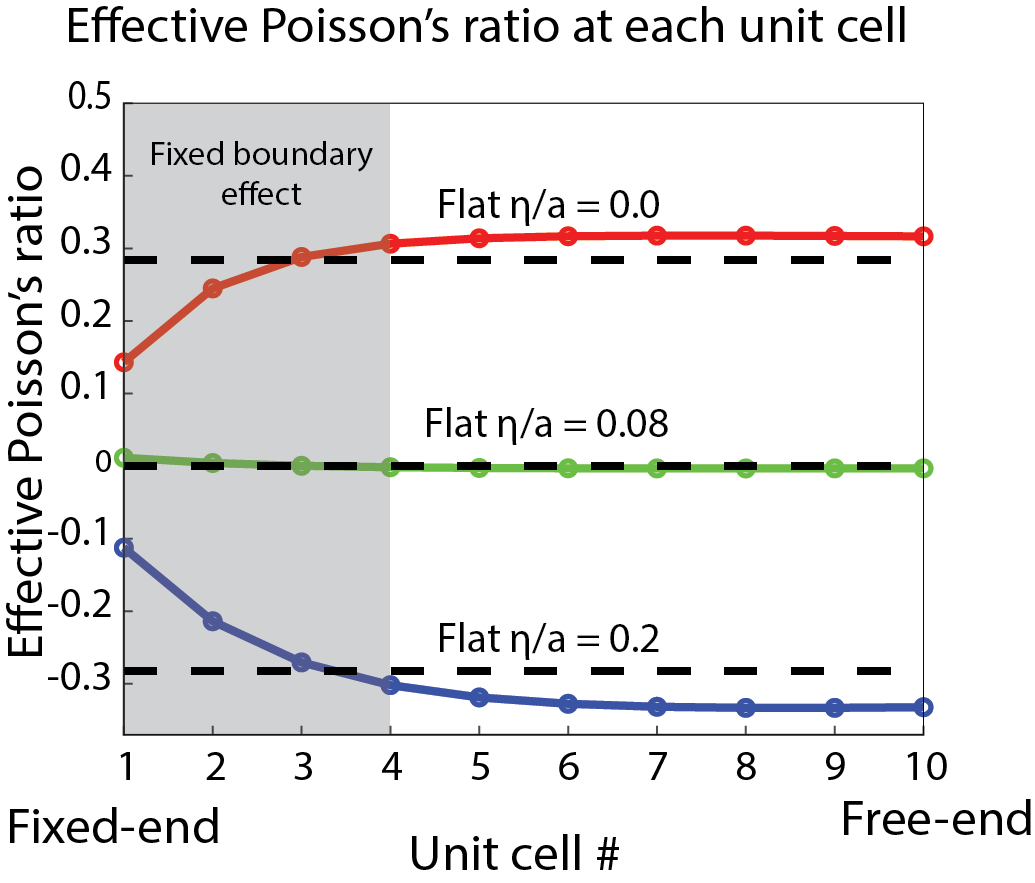}
\caption{\label{fig:S2}\textbf{local Effective Poisson's ratio:} Effective Poisson's ratio at each unit cell for three designs with $\eta / a$ = 0.0, 0.08, and 0.2 with solid lines show local values, and dashed lines show average values.} 
\end{figure}

\end{widetext}
\end{document}